\documentclass[fleqn]{article}
\usepackage{graphicx,amssymb} 
\title{The local deflection of light}
\author{Eduardo D\'{\i}az-Miguel \\
Departamento F\'{\i}sica Aplicada I, Facultad de Ciencias\\ 
Universidad de M\'alaga, 29071- M\'alaga, Spain}

\begin{document}
\maketitle

\begin{abstract}

We have derived the relationship between the radial proper distance, h, and the polar angle, $\varphi$, for a light ray that is emitted and travels in the neighbourhood of the Earth's surface. General relativity predicts that, {\it even locally}, the equation which relates these two physical magnitudes differs from the one stated by the principle of equivalence. More precisely, we have proved that, in the weak field limit, the local physical trajectory, $h=h(\varphi)$, is the one that would correspond to a massive Newtonian particle in a field two times greater. Therefore, and contrary to the generally accepted fact, the principle of equivalence gives only $1/2$ part of the general relativity {\it local} deflection of light.\\

\end{abstract}

 As is well known,   \cite{will}, the classic derivations of the deflection of light
 that only use the principle
of equivalence or the corpuscular theory of light  \cite{eins,sold} yield only $1/2$ part of the general relativity deflection.  The calculation of this bending is based on the fact that the light ray comes in from a great distance, is deflected by the Sun, and then is detected on Earth. This is really a (global) scattering \cite{wein} experiment which detects the general-relativistic curvature of space itself : the in and out trajectories are asymptotically free straight lines. Nevertheless, it is admitted \cite{rindler} that in all theories that accept the principle of equivalence, the {\it local} bending of a light ray would be the one of a Newtonian non-relativistic massive particle. 

   We will establish, using general relativity, that the local trajectory of a light ray does not agree with the mentioned prediction of the principle of equivalence: the trajectory is not the Newtonian one. To this end, we will focus our attention on a photon that is emitted and travels in the neighbourhood of the Earth's surface. The magnitudes that take part in our study are described in Fig. 1: $\varphi$ is the physical angle formed by the radial directions OA and OP,  and h is the proper radial distance between the point P of the photon's trajectory and the Earth's surface. As the photon travels in a weak gravitational field  we must work to first order
 in $\frac{GM}{{c}^2 R} \approx 6.95\,{10}^{-10}$. 
The equation governing the orbit of a photon in Schwarzschild's geometry is  \cite{dinver}

\begin{equation} \label{dife}
\frac{d^{2}u}{d{\varphi}^{2}}+u=\frac{3GM}{c^{2}}u^2 \qquad, \qquad u=\frac{1}{r}
\end{equation}
To first order in $\frac{GM}{{c}^2 R}$, the general solution of Eq. (\ref{dife}) is
   \cite{dinver}
\begin{equation}
u=\frac{\cos\varphi}{D}+\frac{GM}{c^{2}D^{2}}(1+B\sin\varphi+{\sin}^{2}\varphi)
\end{equation} where B and D are two arbitrary constants. We suppose that a photon is shot at the point A of Fig. 1
with an initial direction perpendicular to the Earth's radius. Our initial conditions determine B and D:
\begin{eqnarray}
r(0)=R\Rightarrow \frac{1}{R}=\frac{1}{D}+\frac{GM}{c^{2}D^{2}}\Rightarrow (to\, first\, order)\,R=D-\frac{GM}{c^{2}}\\
r'(0)=0\Rightarrow u'(0)=0\Rightarrow B=0 
\end{eqnarray}

Therefore

\begin{equation}\label{uein}
u(\varphi)=\frac{\cos\varphi}{R}+\frac{GM}{c^{2}R^{2}}(1+{\sin}^{2}\varphi-\cos\varphi)
\end{equation}

 We are interested in the local trajectory in the vicinity of A; so Eq. (\ref{uein}) must  be expanded  to second  order in $\varphi$ :

\begin{equation} \label{erefi}
r(\varphi)=\frac{1}{u(\varphi)}=R(1+\frac{1}{2}(1-3\frac{GM}{c^{2}R}){\varphi}^{2})
\end{equation}

 Now  we want to find the distance, h, between the Earth's surface and the point, P, of the trajectory, measured along the radial direction ($\varphi=constant$) through P. If the spatial metric were Euclidean this distance would be given by $h(\varphi)=r(\varphi)-R$; but the spherical gravitational source warps spacetime in such a manner that, along with  the gravitational time dilation, there is also curving in the spatial radial direction. The principles of general relativity give us the way to calculate the physical distance with the aid of the metric tensor: to first order yet again, the proper radial distance between V and P is given  by

\begin{equation}\label{proper}
h =\int_{R}^{r}\sqrt{g_{rr}}dr=\int_{R}^{r}(1+\frac{GM}{c^{2}r})dr=
(r-R)+\frac{GM}{c^{2}}\ln(\frac{r}{R})
\end{equation}

On the other hand, a central field does not alter the symmetrical disposition of the polar coordinates of the spherical surfaces r=constant. That is to say, in each 2-sphere centred at O (with $r\geq R$), the polar angles have  the same physical meaning as in the flat Minkowskian spacetime. Therefore we substitute  Eq.(\ref{erefi}) into  Eq.(\ref{proper}). The result is 

\begin{equation} \label{eins}
h(\varphi)=\frac{1}{2}R{\varphi}^{2}(1-2\frac{GM}{c^{2}R})
\end{equation}

Let us observe that the radial coordinate R in Schwarzschild spacetime has the property that $R\varphi=s$, where $s$ is the  length of the arc AV along the Earth's circumference. Hence  Eq. (\ref{eins}) can also  be expressed in the alternative form

\begin{equation} \label{einst}
h=\frac{1}{2}s{\varphi}(1-2V_{d})
\end{equation}
 where $V_{d}=\frac{GM}{c^{2}R}$ is the dimensionless Earth surface potential.

 Incidentally, note that the use of the radial proper distance has changed the
$3\frac{GM}{c^{2}R}$ factor of Eq.(\ref{erefi}), which has no direct physical meaning (radial Schwarzschild coordinate), by the (physical)
 $2\frac{GM}{c^{2}R}$ factor of our  result: Eq.(\ref{eins}). I have mentioned this fact because in  a previous article \cite {edu} I erroneously asserted that the correct  factor was $3\frac{GM}{c^{2}R}$. See also the Comment by B. Linet \cite {linet}.

To better appreciate the meaning of Eq.(\ref{eins}), we outline  the  corresponding calculation in the frame  of classical Newtonian mechanics: the  differential Binet-Cauchy equation
of the trajectory 
of a  particle of mass m submitted to the gravitational field $\vec{E_{g}}=\frac{-GM}{r^3}\vec{r}$
is   \cite{gold}

\begin{equation}
\frac{d^{2}u}{d{\varphi}^{2}}+u=\frac{GMm^2}{l^{2}} \qquad, \qquad u=\frac{1}{r}
\end{equation}
where $l$ is the constant angular momentum. If the initial horizontal velocity is $c$, 
then $l=Rmc$. Thus  $\frac{d^{2}u}{d{\varphi}^{2}}+u=\frac{GM}{c^{2}R^{2}}$. The exact solution of the previous linear differential equation (with the same initial conditions: $r(0)=R$ and $r'(0)=0$) is

\begin{equation}
u(\varphi)=\frac{\cos\varphi}{R}+\frac{GM}{c^{2}R^{2}}(1-\cos\varphi)
\end{equation}
 
To first order in $\frac{GM}{c^{2}R}$ and to the second one in $\varphi$, we easily get the Newtonian result:

\begin{equation} \label{new}
h(\varphi)=r(\varphi)-R=\frac{1}{2}R{\varphi}^{2}(1-\frac{GM}{c^{2}R})
\end{equation}

Finally, we must compare  Eqs.(\ref{eins}) and (\ref{new}). The conclusion is unavoidable: the local physical Einsteinian trajectory is the 
one that would be traced by a massive Newtonian particle in a field two times greater.

\begin{figure}[h]

\includegraphics*[width=8cm]{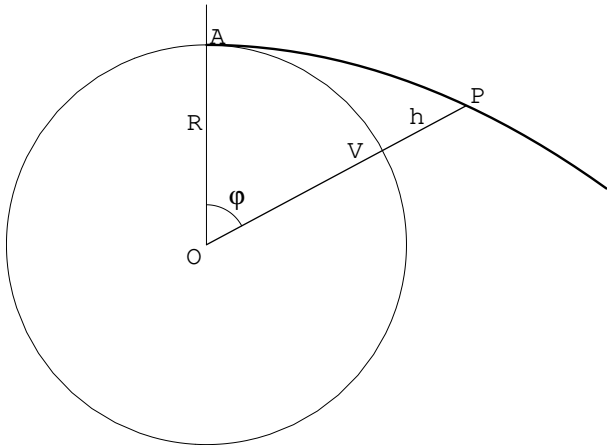}
\caption{\small The deflection of a light ray in the Earth's  gravitational field (greatly exaggerated) 
and the quantities referred to in the text. The Schwarzschild radial coordinate of  P is r.} 
\end{figure}

\end{document}